# Current-driven excitations in magnetic multilayers: a brief review.

**J. Bass, S. Urazhdin, Norman O. Birge,** and **W.P. Pratt Jr.**

Department of Physics and Astronomy, Center for Sensor Materials, and Center for Fundamental Materials Research, Michigan State University, East Lansing, MI 48824

In 1996, Berger and Slonczewski independently predicted that a large enough spin-polarized dc current density sent perpendicularly through a ferromagnetic layer could produce magnetic excitations (spin-waves) or reversal of magnetization (switching). In the past few years, both current-driven switching and current-driven excitation of spin-waves have been observed. The switching is of potential technological interest for direct 'writing' of magnetic random access memory (MRAM) or magnetic media. The spin-wave generation could provide a new source of dc generated microwave radiation. We describe what has been learned experimentally about these two related phenomena, and some models being tested to explain these observations.

## 1 Introduction

Giant Magnetoresistance (GMR) in ferromagnetic/non-magnetic (*F/N*) metallic multilayers is now a major subject in studies of metallic magnetic materials, in part because of its importance for devices such as the read-heads of hard discs. GMR can be described as the change in current passing through the multilayer due to a change in magnetic order—specifically the change in the magnetizations of closest *F*-layers from parallel (*P*) (usually low resistance state) to anti-parallel (*AP*) (usually high resistance state) as an applied magnetic field is reduced from above the saturation field of the *F*-metal to beyond its negative coercive field.

Since its prediction in 1996 [1,2], and supporting evidence of its presence starting in 1998 [3,4], there has arisen great theoretical [5-14] and experimental [15-32] interest in the inverse phenomenon, current-driven excitations in magnetic multilayers: either reversal of layer magnetization, or generation of spin-waves. This interest lies both in trying to understand the physics underlying the new phenomenon and its potential for device use: magnetization reversal for both magnetic memories and magnetic media, and spin-wave generation for production of high frequency radiation. In present magnetic devices or media, moments are reversed via externally generated magnetic fields. It would be much simpler to reverse a moment simply by applying a current pulse perpendicularly through the magnetic layer itself. This possibility is now under study. In this paper, we first describe the most widely used model of current-driven excitations, briefly note others and some issues still to be resolved, and then review what has been learned from experiments.

## 2 Theory.

The most widely used model is the semi-classical spin-torque model of Slonczewski [1], in which a spin-polarized current exerts a torque on an *F*-layer if that layer's moment is not collinear with the direction of current polarization. The current's spin-polarization might be produced, e.g., by passing through another (polarizing) *F*-layer. If the current density is large enough, current in one direction causes the moment of the affected *F*-layer to rotate in one sense, and current in the other direction causes it to rotate in the opposite sense. If the moment of the polarizing layer is fixed in direction, for example by making that layer much thicker or much wider, then the moment of the affected layer can be switched by a large enough current *I* (in a low magnetic field *H*), or set into rotation (in *H* high enough to inhibit reversal). The threshold for excitation is set by competition between the torque and magnetic damping. Models involving a current-induced effective magnetic field [8] or a quantum threshold for excitations [2,3] have also been proposed. Still to be determined by experiment are the relative importances in different circumstances of classical moment rotation versus quantum phenomena, such as incoherent generation of spin-waves. Differences between the two should appear, for example, in the details of thermal activation of magnetization switching at finite temperature. Here, the topic of debate is whether the appropriate temperature is just the phonon temperature, $T_{ph}$, [15,24,28] or an effective current-dependent temperature, $T_m$, [19,30,31]. Experiments must also determine the length scales in the problem [11,12].

## 3 Experiments

The first experimental evidence for a current-driven excitation was obtained at *4.2K* using a *Ag* point contact to a [*Co(1.5) /Cu(2.0-2.2)*] multilayer (all thicknesses in this paper are given in nm) [3]. Since data from that study were included in an MSM02 conference paper two years ago [33], they won't be shown again here, but we just note that peaks in dynamic resistance $dV/dI$ were seen for currents only in one direction, and the current needed to produce a peak grew linearly with the applied magnetic field $H$ (in this case applied perpendicular to the layers). These observations were taken as evidence of excitation of spin-waves. A subsequent experiment [3] showed that microwave radiation applied to such a point contact (by placing the sample and contact within a microwave cavity), gave rise to an additional dc response that was attributed to resonant excitation of spin-waves. Further studies with point contacts at 4.2K showed similar peaks for single *F*-layers [4,16], occasional more complex peak structures [3,4], hysteretic switching [4], current-driven excitations at $H = 0$ for ferromagnetically coupled multilayers [18], and evidence [26] that the critical current at a particular perpendicular field is proportional to the exchange energy density of the *F*-metal (Fig. 1).

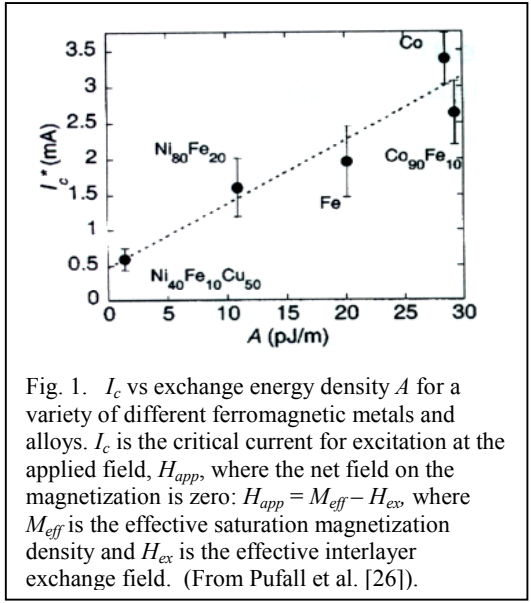

Fig. 1. $I_c$ vs exchange energy density $A$ for a variety of different ferromagnetic metals and alloys. $I_c$ is the critical current for excitation at the applied field, $H_{app}$, where the net field on the magnetization is zero: $H_{app} = M_{eff} - H_{ex}$, where $M_{eff}$ is the effective saturation magnetization density and $H_{ex}$ is the effective interlayer exchange field. (From Pufall et al. [26]).

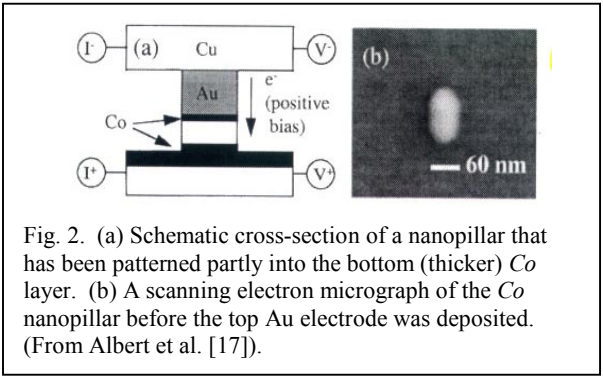

Fig. 2. (a) Schematic cross-section of a nanopillar that has been patterned partly into the bottom (thicker) *Co* layer. (b) A scanning electron micrograph of the *Co* nanopillar before the top Au electrode was deposited. (From Albert et al. [17]).

The first evidence of hysteretic current-driven magnetic switching in a lithographically patterned magnetic nanopillar was reported at room temperature (295K) for *Co/Cu/Co* in [20], and reproduced by others listed above. Switching has also been seen in single and multilayer nanowires electrodeposited into nanochannels [19]. Fig. 2 [17] shows an example of a patterned *Co(40)/Cu(6)/Co(2.5)* nanopillar with dimensions ~ 60 x 130 nm. The elongated shape with sharpened ends is to facilitate formation of only a single magnetic domain. The dipolar coupling between the *Co* layers can be varied by modifying the amount of patterning of the thick bottom *Co* layer. Dipolar coupling is minimized by stopping in the middle of the *Cu* layer, leaving the entire bottom *Co* layer extended. Partial patterning of the bottom layer as in Fig. 2 gives some antiferromagnetic dipolar coupling. Because *Py/Cu/Py* (*Py* = Permalloy = $Ni_{84}Fe_{16}$) often gives stable switching also at 4.2K, it allowed the first comparative studies of switching at *4.2K* and *295K* [30]. Fig. 3 shows data for a *Py(30)/Cu(10)/Py(6)* trilayer with dimensions about 70 nm x 130 nm and patterned to minimize dipolar coupling between the Py layers. The close agreement between the maximum changes in dV/dI with both $H$ and $I$, evidences complete switching. A signature of direct current-driven switching is the asymmetry of the switching, where positive switching current, $I_s^+$, increases $dV/dI$ as the sample switches from a parallel (*P*) orientation of the two *Co* layer magnetizations to an anti-parallel (*AP*) orientation, but negative switching current, $I_s^-$, decreases $dV/dI$ as the sample reverses from *AP* to *P*. In contrast, switching induced by the self-field of the current gives a symmetric pattern vs *I* similar to those of the MR curves vs *H* in the insets to Fig. 3—i.e.

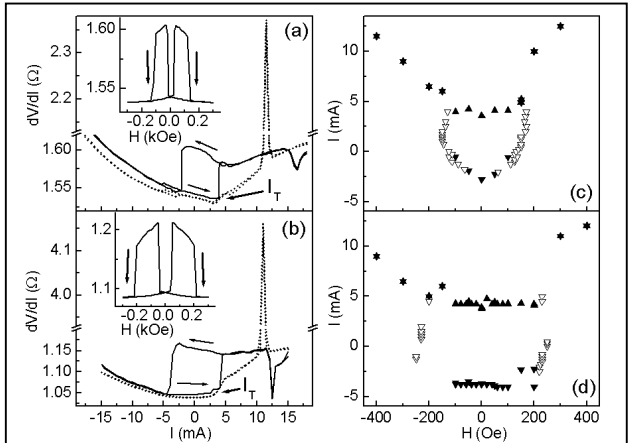

Fig. 3. (a) *295K*; (b) *4.2K*. $dV/dI$ vs $I$ for a nanopillar of *Py(20)/Cu(10)/Py(6)* for $H = 50$ Oe (solid curves) and $H = -500$ Oe (dashed curves). Arrows show scan directions; $I_t$ is the threshold current for a linear rise in $dV/dI$. Insets: $dV/dI$ vs $H$ at $I = 0$. (c) *295K*, (d) *4.2K*: $I$ vs $H$ switching diagrams for the patterned *Py* layer, from current switching at fixed $H$ (solid symbols) or field switching at fixed $I$ (open symbols). Down triangles show *AP* to *P* switching, up *P* to *AP*. Coinciding up and down triangles mark reversible switching peaks. (From Urazhdin et al., [30]).

$dV/dI$ is in the lower *P*-state at both large negative and positive $I$ and in the higher *AP*-state at intermediate $I$.

We now summarize some results of further studies.

We start with the *H-I* 'phase-diagram' for switching in samples with weak coupling between the two *F*-layers. Fig. 3c,d shows that the switching diagrams for *Py/Cu/Py* at *4.2K* and *295K* are qualitatively similar, but much squarer at *4.2K*. As first shown completely in [23], the diagrams are nearly symmetric in $H$ but highly asymmetric in $I$. Beyond a certain value of negative $I^-$, switching disappears. Beyond a certain value of positive $I^+$, switching becomes non-hysteretic (reversible), as shown by a change from a non-reversible step to a reversible peak. Fig. 3a,b shows that, in the reversible regime, $dV/dI$ has a threshold current, labeled $I_t$, above which the data increase almost linearly up to where the sharp peak occurs. As illustrated in Fig. 3a,b, further studies show [30] that $I_t$ is closely equal to the switching current at zero field, $I_s$.

Switching just beyond the hysteretic to non-hysteretic transition was first shown in *Co/Cu/Co* [24] to be associated with telegraph noise switching (see, e.g., Fig. 4a for *Py/Cu/Py*) between the *P* and *AP* states, and the variation of the switching period $\tau$ with $I$ and $H$ was examined at *295K*. More recently [30], the switching peaks in Figs. 3a,b were shown to occur where the dwell times in these two states are approximately equal, $\tau_P \approx \tau_{AP}$. Fig. 4b shows that, when they are equal, this common dwell time (or period) decreases exponentially with $I$, with slopes that are similar at *4.2K* and *295K*. The similarity of these slopes, as well as the detailed variations of $\tau$ with $I$ and $H$ shown in Figs. 4c,d were taken as evidence that the effective temperature for switching is not simply $T_{ph}$ (which would differ by a factor of 70 between *4.2K* and *295K*), but rather a current-driven temperature, $T_m$, that can differ substantially from $T_{ph}$ when the current $I$ exceeds $I_t$.

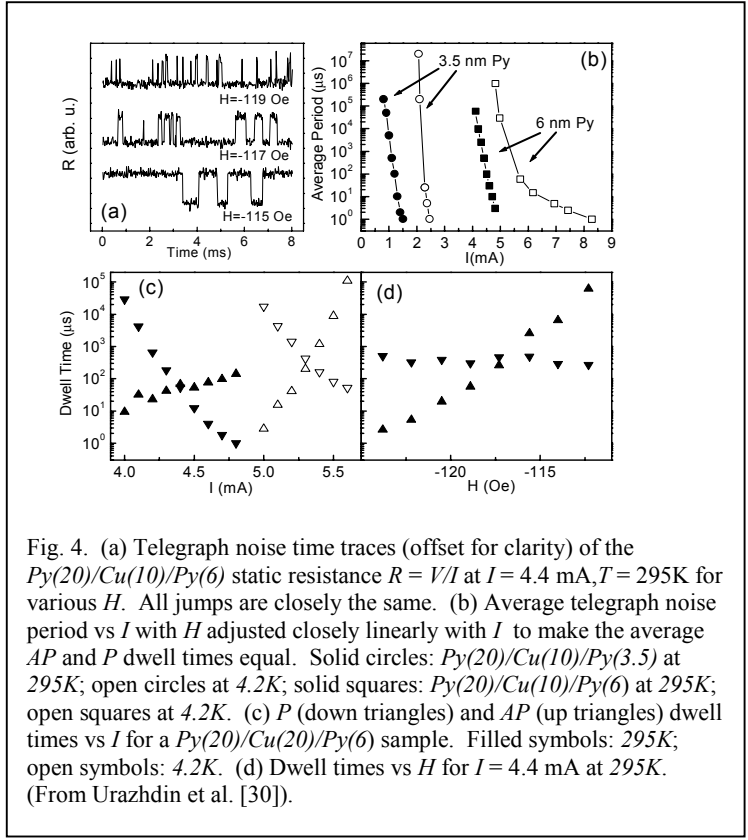

Fig. 4. (a) Telegraph noise time traces (offset for clarity) of the *Py(20)/Cu(10)/Py(6)* static resistance $R = V/I$ at $I = 4.4$ mA, $T = 295K$ for various *H*. All jumps are closely the same. (b) Average telegraph noise period vs $I$ with $H$ adjusted closely linearly with $I$ to make the average *AP* and *P* dwell times equal. Solid circles: *Py(20)/Cu(10)/Py(3.5)* at *295K*; open circles at *4.2K*; solid squares: *Py(20)/Cu(10)/Py(6)* at *295K*; open squares at *4.2K*. (c) *P* (down triangles) and *AP* (up triangles) dwell times vs $I$ for a *Py(20)/Cu(20)/Py(6)* sample. Filled symbols: *295K*; open symbols: *4.2K*. (d) Dwell times vs $H$ for $I = 4.4$ mA at *295K*. (From Urazhdin et al. [30]).

An observation [25] that the switching current, $I_s$, is proportional to the thickness of the free *F*-layer (Fig. 5), was interpreted as evidence of an interfacial source for the current-driven excitations.

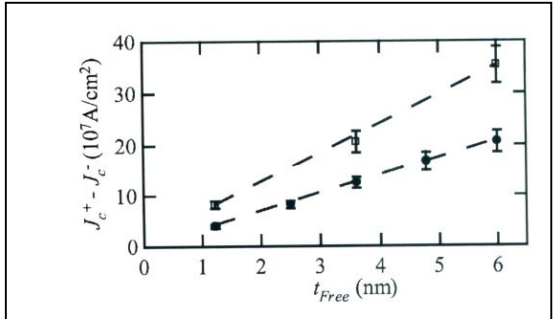

Fig. 5. Critical current density difference, $\Delta J_c = J_c^+ - J_c^-$, at $H = 0$ vs the thickness of the switching $Co$ layer, $t_{free}$, in a $Co/Cu/Co(t_{free})$ nanopillar for current ramp rates of 300 mA/sec (open squares) and 0.1 mA/sec (closed circles). Dashed lines are guides to the eye. (From Albert et al. [25]).

Antiferromagnetic coupling between the two *F*-layers can be produced magnetostatically by at least partly patterning the lower *F*-layer as shown in Fig. 2. Since exchange coupling between two metallic *F*-layers oscillates with the thickness of the spacer layer separating the *F*-layers, ferromagnetic coupling can be induced by leaving the bottom layer unpatterned and choosing a spacer thickness corresponding to ferromagnetic coupling. Fig. 6 [31] shows the different switching behaviors associated with different couplings. Fig 6a-c shows the magnetoresistance (MR) at $H = 0$ and the low field hysteretic and high field reversible switching with telegraph noise expected for an uncoupled sample. Fig. 6d-g shows the MR at $H = 0$ for an antiferromagnetically coupled sample; such a sample can display reversible switching and telegraph noise at low fields. Fig. 6h-j shows that the MR for a ferromagnetically coupled sample can be almost zero; such a sample can display reversible switching and telegraph noise at all fields. These different behaviors can be explained in a simple (but not unique) way using the two-level diagrams of Fig. 7. We start at $H = 0$. In uncoupled samples, the *P* and *AP* wells have the same depth (Fig. 7b) and a large applied current *I* (or $-I$) is needed to increase $T_m^P$ ($T_m^{AP}$) enough to excite the system from either initial state. Antiferromagnetic coupling facilitates the *AP* state, leading to Fig. 7a. Now *P* to *AP* transitions are excited by the ambient temperature, and a large reverse current is needed to produce large enough $T_m^{AP}$ to generate telegraph noise. Conversely, ferromagnetic coupling facilitates the *P* state, leading to Fig. 7c, where a large enough forward current will excite telegraph noise. Turning now to the effect of changing *H*, increasing *H* from $H = 0$ favors the *P*-state. For uncoupled samples, increasing *H* causes Fig. 7b to gradually change to Fig. 7c, shifting the switching from hysteretic to reversible with telegraph noise. For antiferromagnetic coupling, Fig. 7a changes to Fig. 7b and then to Fig. 7c, shifting the switching from reversible at negative I to hysteretic to reversible at positive *I* (Fig. 6e). For ferromagnetic coupling, Fig. 7c stays dominant for all *H*, and the switching is reversible at all positive *I*.

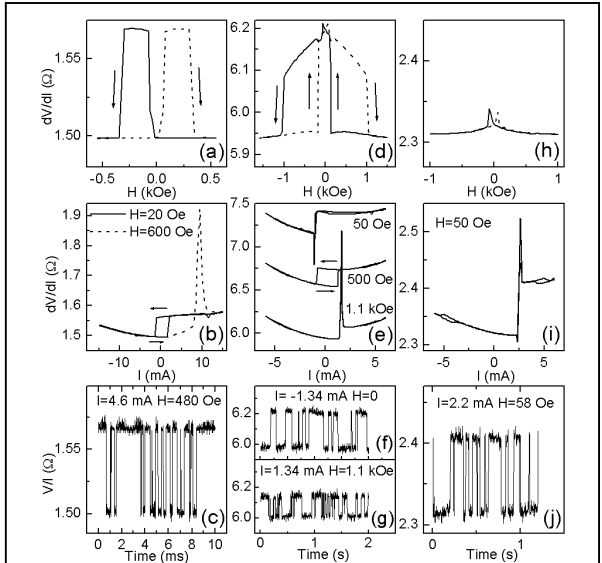

Fig. 6. Magnetoresistance, Current Switching, and Telegraph Noise in uncoupled (a-c), antiferromagnetically coupled (d-g), and ferromagnetically coupled (h-j) trilayer nanopillars. In hysteretic plots, arrows show scan direction. (a,d,h) $dV/dI$ vs $H$ at $I = 0$. (b,e,i) $dV/dI$ vs $I$ at the listed values of $H$. In (e) the curves are offset for clarity. (c,f,g,j) Telegraph noise traces, $R = V/I$ vs time, at the listed values of $H$ and $I$. (From Urazhdin et al. [31]).

As noted above, in a high magnetic field, current-driven excitations in magnetic multilayers excited by a point contact have been seen as peaks in $dV/dI$ and attributed to excitation of spin-waves, i.e. less than full reversal of the moment. In contrast, recent data [27] for nanopillars evidence complete moment reversal. Recent high speed [28] spin-transfer studies were generally consistent with a simple spin-torque model. Detailed high frequency studies are just beginning. One study of nanopillars [29] revealed complex spectra above a minimum current, with both broad backgrounds and peaks that vary in location and relative sizes with current and field. The authors conclude that several different types of magnetic excitations must be present. Another, using a lithographed point contact to a multilayer [32], reported results generally consistent with spin-torque driven excitation of a single domain nanoparticle.

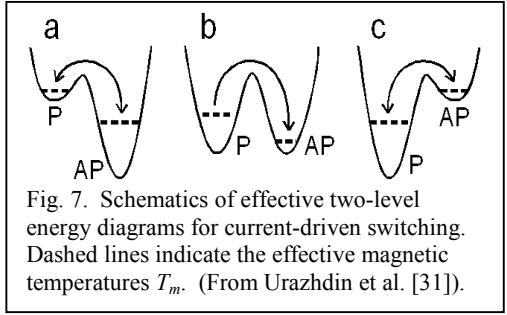

Fig. 7. Schematics of effective two-level energy diagrams for current-driven switching. Dashed lines indicate the effective magnetic temperatures $T_m$. (From Urazhdin et al. [31]).

### 4. Summary and Conclusions

Point contact and nanopillar studies have revealed a fair amount about current-driven excitations in magnetic multilayers. In small fields, the excitation structure of uncoupled layers seems simple—just switching of the magnetization of one layer. Both coupling and higher fields lead to complications, coupling causing reversible switching even at low fields, and higher fields causing more complex excitations, including telegraph noise, spin-waves, and perhaps even chaotic magnetization dynamics. The sources of peaks in point contact spectra are not yet completely clear. Many observed phenomena can be described qualitatively (and sometimes quantitatively—see, e.g. [26,28]), by a simple semi-classical spin-torque model. However, evidence of complications from several experiments suggests that a full understanding of all observations is not yet achieved.

**Acknowledgements** The authors acknowledge support from the MSU CFMR, CSM, the MSU Keck microfabrication facility, the US NSF through grants DMR 02-02476, 98-09688, NSF-EU 00-98803, and Seagate Technology.